\documentclass[iop]{emulateapj}

\shortauthors{Batalha {\em et al.}}

\begin{document}

\title{Challenges to Constraining Exoplanet Masses via Transmission Spectroscopy}

\author{Natasha E. Batalha\altaffilmark{1,2}}

\affil{Department of Astronomy \& Astrophysics, Pennsylvania State University, State College, PA 16802}

\email{neb149@psu.edu}

\author{Eliza M.-R. Kempton}

\affil{Department of Physics, Grinnell College, 1116 8th Avenue, Grinnell, IA 50112, USA}

\author{Rostom Mbarek}

\affil{Department of Astronomy \& Astrophysics, University of Chicago, 5640 South Ellis Avenue, Chicago, IL 60637, USA}

\altaffiltext{1}{Center for Exoplanets and Habitable Worlds, Pennsylvania State University, State College, PA 16802}
\altaffiltext{2}{Planetary Systems Laboratory, Goddard Space Flight Center, Greenbelt, MD 20770}

\begin{abstract}
\textit{MassSpec}, a method for determining the mass of a transiting exoplanet from its transmission spectrum alone, was proposed by \citet{dew13}.  The premise of this method relies on the planet's surface gravity being extracted from the transmission spectrum via its effect on the atmospheric scale height, which in turn determines the strength of absorption features.  Here, we further explore the applicability of \textit{MassSpec} to low-mass exoplanets -- specifically those in the super-Earth size range for which radial velocity determinations of the planetary mass can be extremely challenging and resource intensive.  Determining the masses of these planets is of the utmost importance because their nature is otherwise highly unconstrained.  Without knowledge of the mass, these planets could be rocky, icy, or gas-dominated.  To investigate the effects of planetary mass on transmission spectra, we present simulated observations of super-Earths with atmospheres made up of mixtures of H$_2$O and H$_2$, both with and without clouds.  We model their transmission spectra and run simulations of each planet as it would be observed with \textit{JWST} using the NIRISS, NIRSpec, and MIRI instruments.  We find that significant degeneracies exist between transmission spectra of planets with different masses and compositions, making it impossible to unambiguously determine the planet's mass in many cases.
\end{abstract}

\section{Introduction}
The canonical idea of a “small” planet has dramatically changed in the last decade. Out of the thousands of planet candidates discovered by \emph{Kepler} \citep{mul15}, nearly 80\% of them have radii $<3 R_\earth$. Additionally, occurrence rate studies verify that this high frequency is not merely an observational bias \citep{dre13,pet13,mor14,sil15}. The next step to understanding this large population of planets is to obtain precise mass measurements necessary to shed light on the bulk composition of these objects. Although these masses are unobtainable via \emph{Kepler}'s transit method, they can be calculated from transit timing variations (TTVs) \citep[e.g.][]{lis11,lis13,car12,jon14,mas14,jon16} and radial velocity (RV) measurements \citep[e.g.][]{bat11,wei13,how13,dre15}. Nevertheless, the RV and TTV methods have only yielded masses for a fraction of these systems to date. 

Unlike the low-mass planets in our Solar System, several of these small-radius planets that were expected to be rocky (e.g. the Kepler-11 system, \citet{lis13}), were later found by way of their mass and inferred bulk density to have large H/He envelopes.  Although radius might act as a first order proxy for a planet's composition and H/He fraction \citep{lop14,rog15}, there is no one-to-one mapping between planet radius and mass for sub-gas giant planets.  Considerable compositional degeneracies exist in theoretical mass-radius relations for low-mass exoplanets \citep{for07,sea07}.  Furthermore, observations of the spectra of transiting exoplanets require knowledge of the planet's surface gravity, and therefore its mass, to correctly interpret key properties such as the thermal structure, atmospheric composition, and presence of aerosols. Mass measurements are therefore a necessary first step toward understanding the population of sub-jovian exoplanets. 

Determining masses through RVs (effective for massive planets around bright, quiet stars) and TTV analyses (effective for closely spaced planets in multi-planet systems) is a technical challenge, which in the case of low-mass planets is often insurmountable with current instruments. The current state-of-the art is $\sim$0.8 m s$^{-1}$ while an Earth-mass planet in the habitable zones of a Sun-like star and a 0.1 $M_\sun$ M-dwarf, respectively, will produce a 0.09 m s$^{-1}$ and a 0.9 m s$^{-1}$ signal \citep{fis16}. Although a number of RV instruments able to measure masses down to an Earth-mass and below are currently in development  \citep[see][]{fis16}, most will only begin operation $\sim$1-3 years after the launch of the \emph{James Webb Space Telescope} (\emph{JWST}). Yet a key priority of the \emph{JWST} mission is to characterize the atmospheres of low-mass exoplanets.

In the interest of bypassing resource-intensive RVs to determine exoplanet masses, a technique for determining the mass of a transiting exoplanet from atmospheric observations alone -- via its transmission spectrum -- was proposed by \citet{dew13}.  This method, termed \emph{MassSpec}, relies on accurate determinations of atmospheric temperature, $T$, mean molecular weight (MMW), $\mu$, and scale height, $H= \frac{kT}{\mu g}$, because $M_p=\frac{kTR_p^2}{\mu GH}$. For transiting planets, $R_p$ is known, $T$ is estimated based on the planet's equilibrium temperature, and $H$ is measured from the strength of absorption features in the transmission spectrum.  For hot Jupiters, the MMW is approximately known \emph{a priori} because these planets are assumed to have H/He-dominated atmospheres with $\mu\approx2.3$ amu. In this case, \emph{MassSpec} can be used to verify or determine exoplanet masses.  

Low-mass planets, however, could be rocky, icy, or they could have large H/He envelopes. The implied atmospheric composition that accompanies each type of planet spans a wide range.  This poses a challenge for retrieving masses because $\mu$ is essentially unconstrained. Here, we investigate the extent of these degeneracies and determine the feasibility of extracting masses specifically from \emph{JWST} observations of the transmission spectra of small-radius planets without any \emph{a priori} knowledge of planet mass. In \S2 we describe our method for modeling spectra and \emph{JWST} instrumental noise, in \S3 we describe our results, and we end with concluding remarks in \S4.

\section{Methods}

\subsection{Modeling Transmission Spectra \label{sec:model_spectra}}

In order to investigate degeneracies in transmission spectra for planets of unknown mass, we use the \texttt{Exo-Transmit} radiative transfer package \citep{kem16} to compute a grid of spectra where we vary both the planet's surface gravity and atmospheric composition while keeping the planet's size fixed.  We then inter-compare the spectra to determine which ones are observationally distinguishable from one another.  In all cases, we fix the temperature-pressure profile to an isothermal $T$=400 K and the planet's radius at $R_p = 1.5$  $R_\earth$. This planet size was specifically chosen to reflect the point of greatest compositional uncertainty, which corresponds to the maximum size after which planets tend to decrease in density with increasing radius, indicating a H/He-envelope \citep{wei14,rog15}. In other words, a 1.5 $R_\earth$ planet could be rocky, icy, or could have a large H/He envelope with approximately equal probability. 

For simplicity, we consider atmospheres that are a mix of only H$_2$ and H$_2$O -- two of the major constituent materials for low-mass exoplanets.  This approach is motivated because any absorptive gas will produce the same qualitative behaviors as what we describe in the following sections. Additionally, barring very high C/O planets, self-consistent models that include photochemistry, thermochemistry and kinetic-transport always show large H$_2$O components, regardless of H$_2$ content \citep{hu14}. Therefore, additional molecules are not expected to change our results.

In our model grid, we vary gravity from 5 - 25 m/s$^2$, in steps of 1.4 m/s$^2$, and the ratio of H$_2$O to H$_2$ (by volume) from $10^{-3}$ - $10^{1}$ in log steps of 0.4 dex, creating a total of 150 models. This range of parameters covers H$_2$-rich mini-Neptunes, H$_2$O-rich water worlds, and rocky planets with H$_2$-H$_2$O atmospheres.  We also generate a limited number of spectra for cloudy atmospheres where a fully optically thick gray cloud has been inserted at a specified atmospheric pressure.

Because we only investigate H$_2$-H$_2$O atmospheres, the sole opacity sources in the cloud-free models are the vibration-rotation bands of water vapor in the near- and mid-IR, collision-induced absorption (CIA), and Rayleigh scattering off of H$_2$ and H$_2$O gas (see \citet{fre08, fre14} and \citet{lup14} for a more detailed description of the opacity data).  \texttt{Exo-Transmit} includes H$_2$-H$_2$ CIA opacities but not H$_2$-H$_2$O or H$_2$O-H$_2$O.  The non-inclusion of H$_2$O CIA should not affect our results because these opacities tend to be only weakly wavelength dependent.  For our cloudy models, the cloud opacity is treated as an infinite opacity source at the location of the cloud deck.

\subsection{Modeling JWST Observations and Instrumental Noise \label{sec:noise}}

\begin{figure*}[ht]
\centering
 \includegraphics[angle=90,origin=c,width=.8\linewidth]{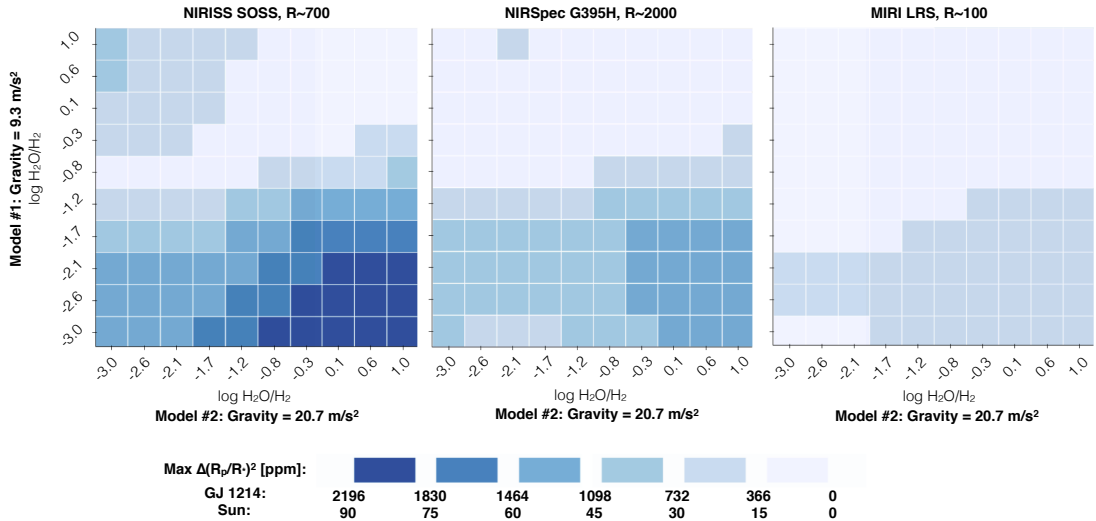}
\caption{Maximum difference between pairs of models binned to the native resolving power of each \emph{JWST} instrument, without instrumental noise.  All models are for a planet with $R_p=1.5$ $R_\earth$ and $T=400$ K. The color scale is indicated for both a GJ 1214-type host star and a Sun-like star. All model \#1's have a surface gravity of 9.3 m/s$^2$, and all model \#2's have a surface gravity of 20.7 m/s$^2$. For reference, the suggested noise floors are $\pm$20 ppm and $\pm$50 ppm for NIRISS/NIRSpec and MIRI LRS, respectively. \label{fig1}}
\end{figure*}

We simulate three different instrument modes: the NIRISS Single Object Slitless Spectrometer (SOSS) (R=700, 0.7-2.7$\micron$), NIRSpec G395H+f290lp (R=2000, 2.8-5$\micron$), and the MIRI Low Resolution Spectrometer (LRS) (R=100, 5-14$\micron$). To calculate the flux and background rates, $F_{\lambda}$ and $B_{\lambda}$, we use the beta version of Space Telescope Science Institute's online Exposure Time Calculator\footnote{https://devjwstetc.stsci.edu} (ETC). The ETC \emph{does not} contain a systematic noise floor, which has been suggested to be anywhere from 20-30 ppm for the near-IR instruments and 50 ppm for MIRI \citep{gre16}. These noise estimates may shift somewhat but will not impact the conclusions in this paper. 

The duty cycle is calculated by determining the number of allowable groups in an integration before detector saturation. A group, in \emph{JWST} terminology, is one or more consecutively read frames --- all exoplanet spectroscopy modes have a single frame per group. To determine the number of groups per integration, $n_{grp}$, we sequentially increase $n_{grp}$ in the ETC, until a single pixel on the detector becomes saturated. The duty cycle is calculated by $d=\frac{n_{grp}-1}{n_{grp}+1}$.  We compute noise simulations for a GJ-1214-like host star with a magnitude $J=8$, as a realistic system that might be observed with \emph{JWST} and also discovered by \emph{TESS} \citep{sul15}. Additionally, a GJ-1214-like star at $J=8$ will have a distance of 6.6 pc, which acts as an optimistic comparison with \citet{dew13}'s assumed system at 15 pc.   
We calculate $n_{grp}= 2$, 9, and 29 for NIRISS SOSS, NIRSpec G395M, and MIRI LRS, respectively, corresponding to duty cycles $d = 0.33$, 0.80, and 0.93. 

Using the duty cycle, the total shot noise is:
\begin{equation}
\sigma_{shot}^2 = F_{in} t_{in}+F_{out} t_{out} 
\end{equation}
$F_{out}$ is the flux (e$^-$/s) for the host star computed from the \emph{JWST} ETC.  $F_{in}$ is the in-transit flux, $F_{in}=F_{out}(1-R_{p,\lambda}^2/R_{*,\lambda}^2)$, where $R_{p,\lambda}^2/R_{*,\lambda}^2$ is obtained from the transmission spectrum model described in $\S$2.1. The in-transit and out-of-transit time components, $t_{in}$ and $t_{out}$, are the transit duration and the out-of-transit observing time, respectively, multiplied by the duty cycle. 

We compute the total noise via
\begin{equation}
    \sigma_{tot}^2 = \frac{1}{(F_{out}t_{out})^2}(\sigma_{shot}^2 + B_\lambda (t_{out}+t_{in}) + RN^2 n_{pix} n_{int})
    \label{eqn2}
\end{equation}
where $n_{int}$ is the total number of integrations during the entire transit, $B_\lambda$ is the extracted background rate (observatory background plus dark current in e$^-$/s), computed in the \emph{JWST} ETC, and $n_{pix}$ is the number of extracted pixels. The read noise, $RN$, will be different for each instrument. We use $RN =18$ e$^-$ for the near-IR detectors and $RN=28$ e$^-$ for MIRI \citep{gre16}. Finally, the factor of $(F_{out}t_{out})^{-2}$ comes from propagating errors from the equation for the transit depth: 
\begin{equation}
z_\lambda = \frac{F_{out}-F_{in}}{F_{out}}.
\end{equation}
The final simulated observation is computed by assuming random Gaussian noise with standard deviation specified by Equation~\ref{eqn2}.  

Finally, unless otherwise specified, each observation simulation is computed for 100 transits. Each transit is composed of 1 hour in-transit, the approximate transit duration for a 400 K, 1.5 R$_\earth$ planet orbiting a GJ-1214-like star, and 1-hour out-of-transit (200 hours for 100 transits). For reference, this is approximately \emph{double} the number of observing hours spent on the longest campaigns for single exoplanet targets to date with \emph{HST} \citep{kre14, kre14b, ste14}.  However, it may be realistic to expect long observing campaigns for the most promising potentially habitable small-radius exoplanets with \emph{JWST}.  The simulated observations from \citet{dew13} are also composed of 200 hours of observing time, but they only consider in-transit data, which are assumed to be spread across three NIRSpec grisms (66.67 hrs in each).  Our observing time is larger by a factor of 3 because we do not assume the time is split between three modes. This only further emphasizes the difficulties in constraining masses via transmission spectroscopy. 

\begin{figure}[ht]
 \includegraphics[angle=0,origin=c,width=\linewidth]{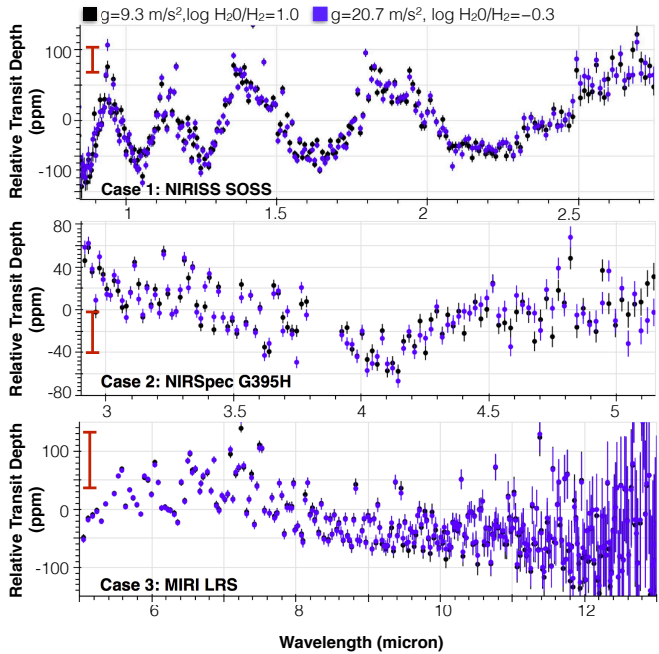}
\caption{JWST simulations for NIRISS SOSS (top), NIRSpec G395H (middle), MIRI LRS (bottom) with no noise floor. Each simulation is done for 100 transits in each observing mode. One transit observation consists of 1 hour in-transit and 1 hour out-of-transit. All simulations are for a GJ 1214-type host star with $J=8$, for the same planet parameters as in Figure~\ref{fig1}. The only difference between the simulations is the gravity and metallicity (H$_2$O/H$_2$), indicated in the color legend. All spectra are binned to $R=100$. The red error bars represent the proposed $\pm$20 ppm noise floor for NIRISS/NIRSpec and $\pm$50 ppm noise floor for MIRI \citep{gre16}.  \label{fig2}}
\end{figure}

\section{Results}

To determine the absolute degree of difference between pairs of model transmission spectra, we directly inter-compare the mean-subtracted spectra at native instrument resolving power without any added noise to find the maximum deviation at a single wavelength. For this comparison of the spectral models, we produce our full model grid (as described in Section~\ref{sec:model_spectra}) for both a Sun-like and a GJ 1214b-like (0.2 $R_{\odot}$) host star.  We find that in each of the observing modes, for the larger Sun-like host star, over 70\% of the planet-pair scenarios have maximum spectral differences of less than 50 ppm. For reference, this is the noise floor that has been suggested for the mid-IR \emph{JWST} instruments \citep{gre16} and is close to the minimum noise level achieved with near-IR detectors on \emph{HST} \citep{lin16}.  That is to say, that in over 70\% of cases, the maximum difference between pairs of models is small enough that it would be challenging for any amount of \emph{JWST} observing time to reveal the distinction. This high percentage is expected because a 400 K super-Earth orbiting a Sun-like star is a challenging target for \emph{JWST} with its long orbital period and small transit depth of approximately 200 ppm. Super-Earths with temperatures of $\sim$400 K around M-dwarf stars, however, have been shown to be attainable for atmospheric characterization with JWST with $\sim$20 transits \citep{bat15,bar15,gre16}. For a GJ 1214-like star, we find 4.5\% (505 total pairs) of cases to be degenerate with maximum differences of 50 ppm or less in the NIRISS band, 5.5\% (615 total pairs) in the NIRSpec G395H band and 13\% (1471 total pairs) in the MIRI LRS band. As expected, these numbers are dependent on the bin size. For example, if a NIRISS observation is binned down from its native resolving power ($R=700$) to $R=100$, the number of degenerate planet pairs at the 50 ppm level increases from 4.5\% to 9\% (1086 total pairs).

In Figure~\ref{fig1} we illustrate the degenerate parameter space by isolating two representative gravities (9.3 and 20.7 m/s$^2$) and showing the maximum spectral differences as a function of composition, with scaling for both the GJ 1214-type and Sun-like host star.  We choose these two gravities to illustrate cases where degeneracies would exist, yet the implied planet types are very different. The planet denoted as model \#1 has a density of 3.5 g/cm$^3$ as opposed to 7.7 g/cm$^3$ for model \#2. By comparing against theoretical mass-radius relationships for low-mass planets \citep{for07,sea07}, the former is a low-density planet with a considerable volatile component -- either an ice-rich water-world or a hydrogen-rich mini-Neptune -- whereas the latter is a rocky planet consistent with an Earth-like bulk composition.  

The least degenerate region of Figure~\ref{fig1} lies in the lower right-hand corner where the high surface gravity planet combined with a high MMW atmosphere will produce much smaller spectral features than a low surface gravity, low MMW atmosphere.  The most degenerate regions occur along a diagonal that cuts across the top portion of each panel of Figure~\ref{fig1}.  This is the region where both model \#1 and model \#2 produce comparably sized spectral features, which takes place where the low surface gravity planet has a higher MMW atmosphere and vice versa.  This qualitative behavior extends to other surface gravity pairings.  We point out that the relative strength of spectral features in transmission grows proportionately with temperature, as will the maximum spectral differences between models. Planets with $T>400$~K will therefore be more easily distinguishable from one another, and less so for $T<400$~K.  
 
The metric employed in Figure~\ref{fig1} for distinguishing between atmospheric scenarios assumes that model discrepancies at a single wavelength are sufficient for ascertaining the best-fit model parameters.  More realistically, instrumental noise along with finite observing time will limit the degree to which an atmosphere can be characterized.  To quantify these effects, in Figure~\ref{fig2}, we isolate a single degenerate planet pair from Figure~\ref{fig1} and show each resulting simulated observation in the three different \emph{JWST} modes using the noise model described in Section~\ref{sec:noise}.  The two planets shown in Figure~\ref{fig2} have $g = 9.3$ m/s$^2$ with $\log[$H$_2$O/H$_2]=1.0$ and $g = 20.7$ m/s$^2$ with $\log[$H$_2$O/H$_2]=-0.3$. The former case represents a water world with a water-dominated atmosphere, whereas the latter would be a rocky planet with an outgassed hydrogen-dominated atmosphere.  

On average over each observational band pass, the difference between the two observations is $\sim$10 ppm. For NIRISS SOSS (the least degenerate observation), only considering pure shot and read noise, the errors after 100 transits are $\sim$5 ppm. The reduced-$\chi^2$ for one NIRISS observation using the opposing model as a template is $\chi_{red}^2=1.3$ --- right at the 3-$\sigma$ interval for 230 DOF ($\chi_{red}^2< 1.27$) it would just be on the cusp of non-degeneracy. In reality though, 5 ppm error bars are highly unlikely given the current state of the art using \emph{Hubble} \citep[e.g.][]{lin16}, as well as current knowledge of instrument systematics \citep{gre16}. If we were to include a 20 ppm noise floor, the reduced-$\chi^2$ shrinks to $\chi_{red}^2=1.06$. Within the 1-$\sigma$ interval for 230 DOF ($\chi_{red}^2< 1.09$), these cases could only be distinguished with \emph{a priori} knowledge of the planet's mass.  
\begin{figure}[ht]
 \includegraphics[angle=0,origin=c,width=\linewidth]{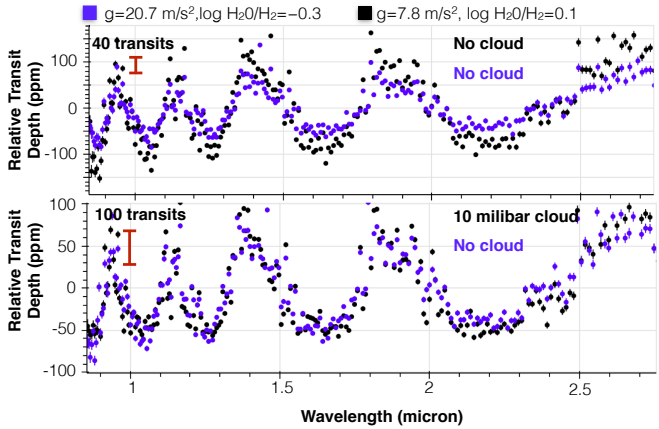}
\caption{Simulated data for a GJ 1214-type system for the same stellar and planetary parameters as the previous two figures. The two simulations differ in their assumed gravity, metallicity (H$_2$O/H$_2$), indicated in the color legend, and cloud assumptions. In the top panel neither model has clouds. In the bottom panel, the low gravity, high metallicity model has a 10 millibar cloud that reduces the strength of its absorption features. Observations were simulated for NIRISS SOSS with no noise floor and binned to R=100. Top panel observations were simulated with 40 transits (80 hrs). Bottom were simulated with 100 transits (200 hrs). Some error bars are too small to see.  The red error bars represent the proposed $\pm$20 ppm noise floor for NIRISS \citep{gre16}. \label{fig3}}
\end{figure}
Given the considerable number of non-degenerate planet pairs, especially for the M-dwarf host star, it may seem that \textit{MassSpec} could be a productive method for measuring exoplanet masses in some situations where RVs are unattainable.  However, the aforementioned cases were all modeled without the presence of aerosols (clouds and haze). As we know from ground and space-based observations, aerosols are practically ubiquitous in the transmission spectra of exoplanets \citep[e.g.][]{kre14,sin16}.  Aerosols add an additional layer of degeneracy to the retrieval of physical parameters from transmission spectra and can therefore further complicate the extraction of mass information from atmospheric observations.  This occurs because both the presence of clouds and an increase in metallicity can have the same dampening effect on molecular absorption features.  In Figure~\ref{fig3}, we demonstrate this effect by adding a gray opacity source to one of two planet cases whose cloud-free spectra were initially non-degenerate. We show that when a 10 mbar cloud is added to a target with $g = 7.8$ m/s$^2$ and $\log[$H$_2$O/H$_2]=0.1$ it becomes degenerate with a target that has a no clouds, $g=20.7$ m/s$^2$, and $\log[$H$_2$O/H$_2]=-0.3$. Figure~\ref{fig3} is a single illustration of a general effect, in which we expect the presence of aerosols to dramatically increase the number of planet pairs that are degenerate with one another.     

\section{Discussion \& Conclusion}

To investigate degeneracies between gravity and composition in small planets, we have inter-compared 150 forward model transmission spectra for a planet with a H$_2$-H$_2$O atmosphere, radius of $R=1.5$ R$_\earth$, and isothermal temperature of 400 K. The surface gravity and atmospheric composition were allowed to vary with the goal of determining whether the planet's mass (via its surface gravity) is recoverable from transmission spectrum observations.  We found that a considerable fraction of the planet pairs were identical to one another at or below the 50 ppm level --- more so for a larger Sun-like host star.  With the addition of clouds, these degeneracies were exacerbated.  The 50 ppm level is important because it is the approximate minimum noise level, or noise floor, that has been suggested for mid-IR \emph{JWST} instruments \citep{gre16}. Barring any unforeseen circumstances, the near-IR instruments will likely have a lower noise floor of 20-30 ppm. 

By modeling the \emph{JWST} noise sources we determined that even 100 transits (equivalent here to 200 hrs) in key observing modes is not sufficient to discern between many planet pairs, even assuming no systematic noise floor. A shorter timespan of observations, smaller planetary radius, larger stellar radius, or lower planetary temperature will further enhance the difficulties with extracting a planet's mass from its transmission spectrum.

These conclusions paint a far more pessimistic picture of mass extraction via transmission spectroscopy than \citet{dew13}. Yet our results are fully consistent with retrieval studies \citep{ben12,ben13,bar15,gre16}, which attempt to constrain the atmospheres of super-Earths and mini-Neptunes with simulated \emph{JWST} data, even when the masses \emph{are known}. For example, \citet{gre16} find that only a $\log[$H$_2$O$]>-7$ lower limit can be placed on the water mixing ratio of a cloud-free, 100\% H$_2$O, 500~K, 2.1~R$_\earth$ planet orbiting an K=8 M0.0V star with a high SNR 1-11~$\micron$ observation. Similarly, \citet{ben12} cannot reliably determine whether the observed absorber is the main constituent of the atmosphere or just a minor species when the mixing ratio is less than 0.1\%, and there is no observation of the molecular Rayleigh scattering ($<1 \micron$). We speculate that \citet{dew13}'s optimistic results may be in part because of narrow bounds on priors, assumptions of higher \emph{JWST} duty cycles, their modeling of only in-transit observations, and/or their lack of systematic noise. 

Our choice to model a 1.5 $R_{\earth}$ planet was motivated by the observation that planets of this size have bulk densities that are equally likely to indicate a rocky planet as one that is volatile-rich \citep{lop14,rog15}.  These two planet types have very different implied formation histories, with the latter more likely to have formed beyond the snow line and migrated inward.  Yet we have shown a case where a water world is indistinguishable from a high surface gravity rocky planet with \emph{JWST} observations (Figure~\ref{fig2}), and many more similar degenerate planet pairs exist in our full grid.  Therefore, independently determining the mass of small-radius planets is required in order to correctly interpret the composition of a planet's atmosphere and its formation history.

We caution that \emph{JWST} observations undertaken for atmospheric characterization of small-radius planets whose masses are unknown may not yield fruitful results.  Further down the road, spectroscopic characterization efforts for directly-imaged near Earth-size planets are also likely to present substantial model degeneracies, absent mass measurements.  In the \emph{JWST} era and beyond, successful interpretation of exoplanet spectra for small-radius planets will therefore rely necessarily on the success of future precision RV instruments such as Carmenes, HPF, MAROON-X, NEID, Spirou, and Veloce.

\acknowledgments
This research has made use of the NASA Exoplanet Archive, which is operated by the California Institute of Technology, under contract with the National Aeronautics and Space Administration under the Exoplanet Exploration Program. This material is based upon work supported by the National Science Foundation under Grant No. DGE1255832 and the Kavli Summer Program in Astrophysics to N.E.B. Any opinions, findings, and conclusions or recommendations expressed in this material are those of the author(s) and do not necessarily reflect the views of the National Science Foundation.  E.M.-R.K. received support for this work from the Research Corporation for Science Advancement via the Cottrell Scholar program and from the Grinnell College Harris Faculty Fellowship.  R.M. was supported by the Grinnell College Mentored Advanced Project (MAP) program.

\end{document}